\pdfoutput=1
\documentclass[12pt,preprint]{aastex}
\usepackage{graphicx}

\newcommand{\et}{et al.}
\newcommand{\kms}{km s$^{-1}$}
\newcommand{\ha}{H$\alpha$}
\newcommand{\solar}{\ifmmode_{\sun}\else$_{\sun}$\fi}

\newcommand{\HI}{H$\,${\sc i}}

\newcommand{\coldens}{atoms cm$^{-2}$}

\begin{document}

\title{Mapping the Extended \HI\ Distribution of Three Dwarf Galaxies}

\author{Deidre A. Hunter\footnotemark[1], 
Fakhri Zahedy\footnotemark[2],
Emily C. Bowsher\footnotemark[1]~~\footnotemark[3]
}
\affil{Lowell Observatory, 1400 West Mars Hill Road, Flagstaff, Arizona 
86001 USA}
\email{dah@lowell.edu, fsz@mit.edu, bowsher@chara.gsu.edu}

\author{Eric M. Wilcots\footnotemark[1], 
Amanda A. Kepley\footnotemark[1]~~\footnotemark[4]
and Veronika Gaal\footnotemark[1]}
\affil{Department of Astronomy, University of Wisconsin, 475 N.\ Charter St.,
Madison, Wisconsin 53706-1582 USA} 
\email{ewilcots@astro.wisc.edu, aak8t@mail.astro.virginia.edu}

\footnotetext[1]{\rm Guest Observer, Green Bank Telescope, National Radio Astronomy Observatory,
a facility of the National Science Foundation operated under cooperative
agreement by Associated Universities, Inc.}
\footnotetext[2]{\rm  Current address:  Department of Physics,
MIT, 77 Massachusetts Avenue, Cambridge, Massachusetts 02139-4307}
\footnotetext[3]{\rm  Current address: Center for Astrophysics and Space Sciences, 
University of California San Diego, 9500 Gilman Drive, La Jolla, California 92093-0354
and Columbia Astrophysics Laboratory, Columbia University, Mail Code 5247, 550 West 120th Street,
New York, New York 10027}
\footnotetext[4]{\rm  Current address: Department of
Astronomy, University of Virginia, Charlottesville, Virginia 22904-4325}

\begin{abstract}
We present large field \HI-line emission maps obtained with the
single-dish Green Bank Telescope
centered on the dwarf irregular galaxies Sextans A, NGC 2366, and WLM.  
We do not detect the extended skirts of emission associated with
the galaxies that were reported from Effelsberg observations (Huchtmeier et al.\ 1981).
The ratio of \HI\ at $10^{19}$ \coldens\ to optical extents of these galaxies are instead 
2--3, which is normal for this type of galaxy.
There is no evidence for a truncation in the \HI\ distribution $\geq10^{19}$ \coldens.
\end{abstract}


\section{Introduction} \label{sec-intro}

Dwarf irregular (dIm) galaxies, like many disk systems, often have neutral
atomic hydrogen gas that extends beyond, and sometimes well beyond,
the optical galaxy. In extreme cases this gas is known to extend
3--7 times farther than the stellar distributions (Huchtmeier \et\ 1981, Bajaja \et\ 1994).
Such envelopes were believed to be relatively quiescent
reservoirs of gas, as there was little data to the contrary.  This changed with Very Large Array 
(VLA\footnotemark[5])
\footnotetext[5]{\rm
The VLA and GBT are facilities of the National Radio Astronomy Observatory.
NRAO is a facility of the National Science Foundation operated under cooperative
agreement by Associated Universities, Inc.}
mosaics of NGC 4449 (Hunter \et\ 1998) and IC 10
(Shostak \& Skillman 1989, Wilcots \& Miller 1999).
NGC 4449 is surrounded
by a complex network of streamers that represents the remains of an
\HI\ disk disrupted by an encounter with another galaxy
(Hunter \et\ 1998, Theis \& Kohle 1998).  IC 10 is embedded in a
counter-rotating halo of \HI\ and also appears to be currently accreting
a large cloud of gas (Wilcots \& Miller 1998).
On the other hand, some dwarfs {\it are} embedded
in large, quiescent disks that are simply smooth extensions of the
galaxy that happen to stretch some 5--7 times farther than the stars
(Carignan \& Purton 1998; Kreckel \et\ 2011). 

We do not yet understand how some irregulars have maintained
extremely large quiescent disks, especially given the onslaught of
activity and ionization in the early universe (Ricotti \et\ 2001).
Similarly, it is not yet clear
what role the extended gas around irregulars plays in 
the evolution of the stellar disk.  
Also, the extended gas is more sensitive to interactions with other galaxies.
Therefore, it is of interest
to determine the full extent of \HI\ distributions around galaxies.

One galaxy with a potentially large smooth extension of gas 
is the dwarf irregular Sextans A. 
Single dish maps of Sextans A by Huchtmeier \et\ (1981) with the Effelsberg 100 m
telescope showed \HI\ extending to a diameter of 54\arcmin\
at a column density of 10$^{19}$ \coldens.
Thus, \HI\ was detected to 8.7 times the galaxy's optically-defined Holmberg diameter.
This motivated 
Wilcots \& Hunter (2002) to map Sextans A in \HI\ using the VLA
in a mosaic that covered 90\arcmin\ diameter at 1\arcmin\ resolution (see also, Barnes \& De Blok 2004).
Although those observations reached a column density of 
$7.5\times10^{18}$ \coldens, \HI\ was only detected to a diameter
of 18\arcmin, about one-third of the extent reported by 
Huchtmeier \et, and only about 62\% of the total \HI\ flux reported
by Huchtmeier \et\ was recovered.
Wilcots and Hunter suggested that the missing gas exists in a very
smooth, low column density distribution with a size scale too
large to be resolved with the interferometer.
They estimated that the density fluctuations in this extended gas
should be no smaller than 10\arcmin--15\arcmin\ to be hidden from 
the VLA.

If the extended gas around Sextans A is indeed that smooth,
it is a prime target for mapping with the 
Green Bank Telescope (GBT\footnotemark[5]).
First, because of its unique design, the GBT has extraordinarily low side lobes:
30 dB below the primary beam at 1420 MHz, about 0.1\%.
This allows one to detect faint gas far from the center of the galaxy.
Second, with a beam-size at 21-cm of 8.7\arcmin\ the GBT would still be able
to resolve large-scale structure in the extended gas if it exists.
We, therefore, undertook to use the GBT to map the \HI\ gas around
Sextans A. Our purpose was to use the unparalleled sensitivity
of GBT to confirm the existence of the 
gas at the outer edges of Sextans A, determine any structure in
the gas, and explore how far it extends and how the density falls off
with radius. 

Two other dwarf galaxies---WLM and NGC 2366---have 
also been reported by Huchtmeier \et\ (1981) 
to have extended \HI\ with diameters that are 3-3.6 times their
optical Holmberg diameters, and this gas has gone undetected by others
(Hunter \et\ 2001, Barnes \& De Blok 2004, Jackson \et\ 2004).
Both are fairly typical nearby, gas-rich dIm galaxies with modest absolute magnitudes.
These two galaxies were also included in this study. 
Thus, we have observed three dIm galaxies with potentially extended \HI\ 
disks---Sextans A, WLM, and NGC 2366---with
the GBT, and we report the results of those observations here.
Basic properties of the galaxy sample are given in Table \ref{tab-sample}.

\section{Observations and Mapping} \label{sec-obs}

We used the 100 m single-dish GBT 
with the GBT Spectrometer 28--29 June 2003
to map the 21-cm \HI-line emission associated with our three galaxies. 
The spectrometer was used in its auto-correlation total-power mode
with a 12.5 MHz bandwidth and 9-level sampling mode
to yield a channel separation of 0.32 \kms\ and a full spectral coverage of 2595 \kms. 
Our central velocity was the heliocentric radial velocity
of the galaxy, given in Table \ref{tab-obs}.
We observed 3C48 at the beginning of the observing session in order to verify pointing. 

The beam-size at our frequencies is about 8.72\arcmin, and for most of the mapping we stepped
every 2.9\arcmin, 1/3 of a beam-size, in order to produce a fully
sampled map. One exception was a portion of the NGC 2366 map to the north that was 
begun with a step of a 1/2 beam-size.
The other exception was the map of WLM which consists of an east-west strip 3 pointings wide 
(one offset 1/3 beam width to the north and one offset 6\arcsec\ to the south) 
and a north-south strip through the galaxy center.
The map center, number of pointings, and final field of view of each map
is given in Table \ref{tab-obs}.
We began the map in the center row and worked our way north one row at a time, 
and then returned to the center and worked our way south.
For Sextans A, the last row of the northern part of the map was only partially
completed due to time limitations.
We used position switching for calibration and background subtraction, and our reference position
was located 40-50\arcmin\ north or south, whichever was closer.
We moved to this off position every 5 steps along the map for Sextans A and WLM 
and every 6 steps for NGC 2366.
We integrated for 1 minute at each position, obtaining 3 spectra of 20 s each and 
expecting to reach an RMS of 0.03 K. 

The data were reduced using GBTIDL.
Because we were interested only in a relatively narrow velocity range around
the central velocity of the galaxies, we eliminated approximately the first and last 1000 channels
in every scan in order to remove noise spikes. 
We calibrated each row pointing against the
reference scan taken during the observations of that row. 
The resulting spectra showed stable, flat baselines. 
A few scans had curvy baselines, and those scans were flagged and not used further.
The continuum on each
side of the emission line was fit and subtracted to produce a baseline at zero.
We smoothed along the velocity axis with a boxcar average of 5 pixels
and proportionally decreased the number of channels in the final spectra.
The final spectra contained 1201 channels for Sextans A and NGC 2366 and 1361 channels
for WLM with channel widths of 1.61 km s$^{-1}$.
We used a standalone program, idlToSdfits, to convert the calibrated spectra
in GBTIDL to a format that mimics radio interferometry uv data and then read
those data into AIPS. Within AIPS we used SDGRD to produce map cubes
with 174\arcsec\ pixels. 

From the map cubes we made moment zero maps of integrated \HI\ emission. 
We integrated over channels 516--579 for Sextans A, 506--594 for NGC 2366,
and 610--677 for WLM.
In the calibration process in GBTIDL, the GBT data are converted to 
$T_a^*$, antenna temperature corrected for atmospheric attenuation,
radiative loss, and rearward scattering and spillover. 
To convert to $T_b$, the brightness temperature of the source, we correct
for the forward scattering and spillover, $\eta_{fss}$, and for the main beam
efficiency, $\eta_{mb}$.  Here we ignore forward scattering and use a typical value
for $\eta_{mb}$ at 1.4 GHz on the GBT of 0.94 so $T_b = 1.07\times T_a^*$
(Braatz 2009).
Then for our beam-size and frequency, $S {\rm (Jy)} = T_a^*/2.02$.
Observations of the standard continuum source 3C48 yield a flux that is 6\% low. 
However, this is within the uncertainty of
10\% for 21-cm calibration reported by Robishaw \& Heiles (2009).
Thus, we expect our fluxes to be uncertain at the 10\% level.

In the map cubes, the units of K beam$^{-1}$ pixel$^{-1}$ can be converted
to units of Jy by multiplying by a factor of (pixel scale)$^2$/beam-size$^{2}$/2.02,
which is a factor of 0.054.
For contours on the moment zero (integrated \HI) maps, map units of
15.3 K beam$^{-1}$ km s$^{-1}$ pixel$^{-1}$ correspond to a column density 
of $10^{19}$ atoms cm$^{-2}$.

\begin{figure}
\includegraphics[angle=0,width=1.0\textwidth]{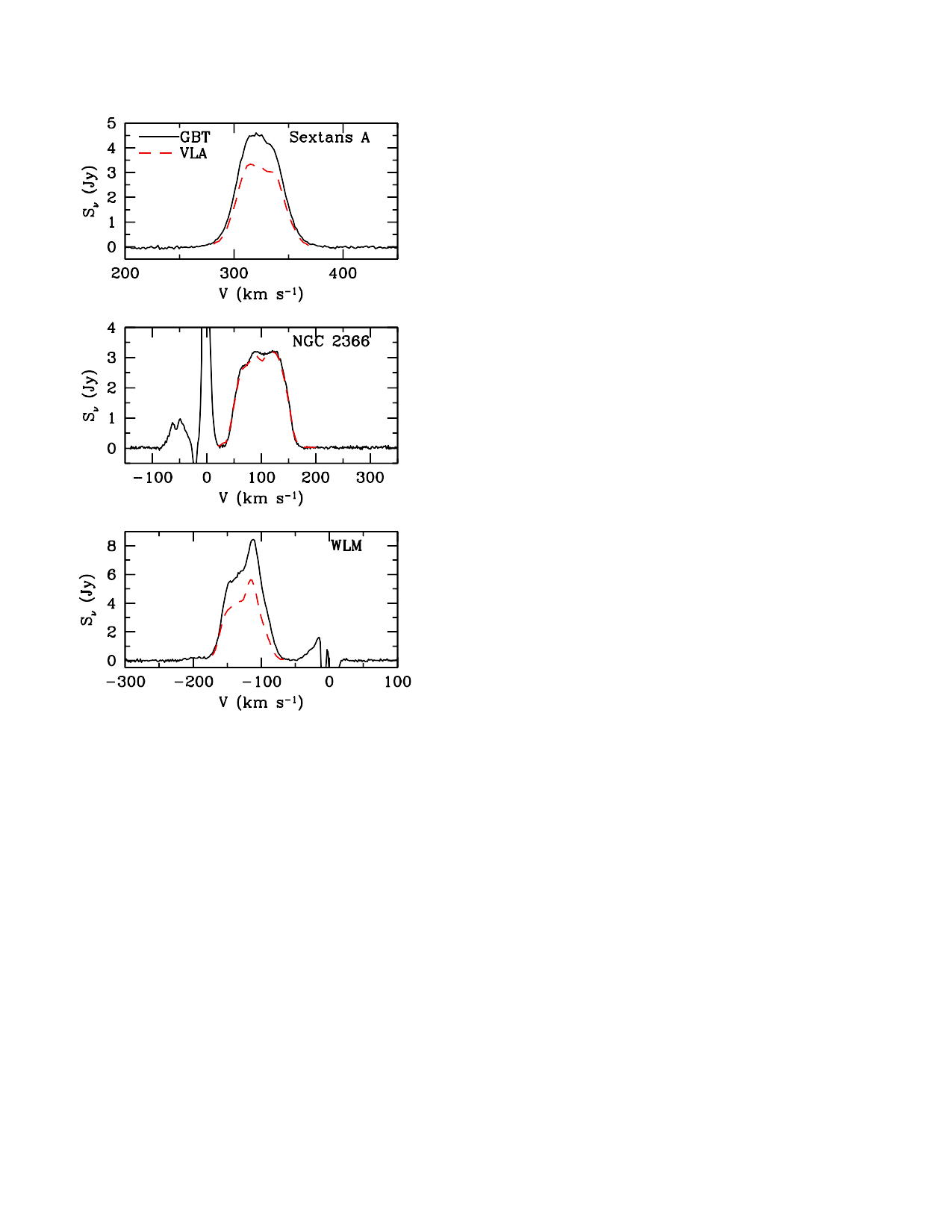}
\caption{Integrated flux in individual channel maps plotted against channel velocity. 
Emission around zero in the NGC 2366 and WLM plots is due to the Milky Way.
Profiles from VLA observations are shown as dashed lines: 
Sextans A data are from Wilcots \& Hunter (2002),
NGC 2366 data are from Hunter \et\ (2001),
and WLM data are from Kepley \et\  (2007).
In spite of known inaccuracies in 21-cm line profiles measured with the GBT
(Robishaw \& Heiles 2009), the GBT profiles agree well with those measured with the VLA.
\label{fig-intprofiles}}
\end{figure}

Our flux-velocity profiles are shown in Figure \ref{fig-intprofiles}.
Robishaw \& Heiles (2009) have reported inaccuracies
in 21-cm line profiles measured with the GBT. 
They find 10\% errors in the calibrated gain and some contributions from distant sidelobes.
Although these problems may increase the uncertainties,
the profile shapes agree well with those measured with the VLA.

\section{Results} \label{sec-results}

\subsection{The Extent of \HI}

\subsubsection{Sextans A}

We detected \HI\ emission from Sextans A between 270 \kms\ to 372 \kms.
This is very close to the range of velocities
detected by Wilcots \& Hunter (2002) and Huchtmeier \et\ (1981)
although their channel spacings of
2.6 \kms\ and 6.6 \kms, respectively, were coarser than ours. 
A Galactic High Velocity Cloud (HVC), identified by Huchtmeier \et, is also
present in the direction of Sextans A. 
We detect it from 110 \kms\ to 152 \kms.

\begin{figure}
\includegraphics[angle=0,scale=0.9,width=1.0\textwidth]{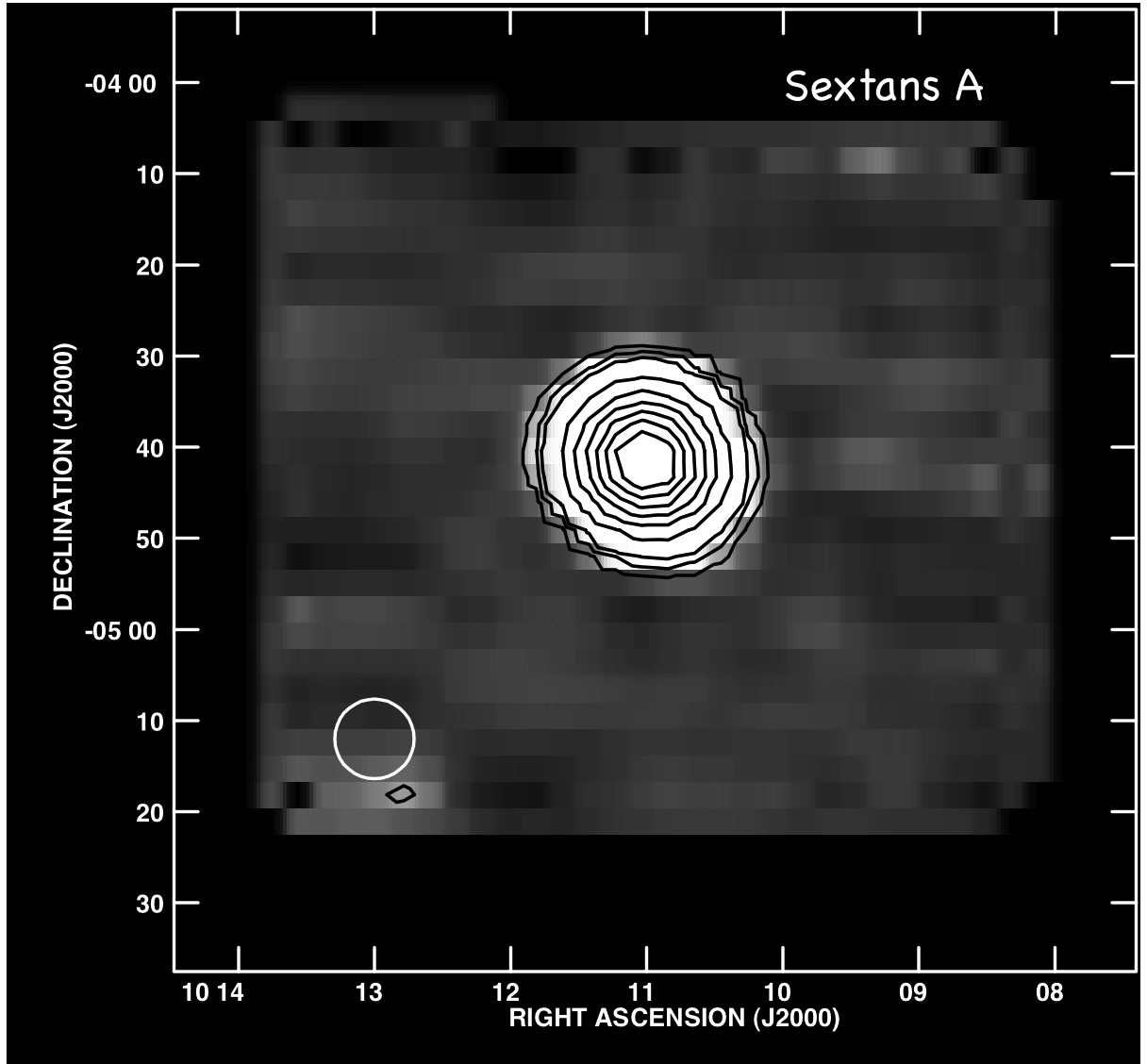}
\caption{Integrated \HI\ map of Sextans A summed from 270.5 \protect\kms\
to 371.9 \protect\kms\ in the GBT data cube. The image is displayed
to bring out the fainter outlying structure.
The circle in the bottom left corner is the FWHM of the beam: 8.72\arcmin\ diameter.
Map pixels are 2.9\arcmin, 1/3 of a beam-size.
Contours are 5, 10, 15.3, 40, 70, 100, 130, 160, and 190 K beam$^{-1}$ km s$^{-1}$ pixel$^{-1}$.
The third contour corresponds to a column density of $10^{19}$ cm$^{-2}$, the isodensity contour
to which we have measured the HI diameters.
\label{fig-sexam0}}
\end{figure}

\begin{figure}
\includegraphics[angle=0,scale=0.9,width=1.0\textwidth]{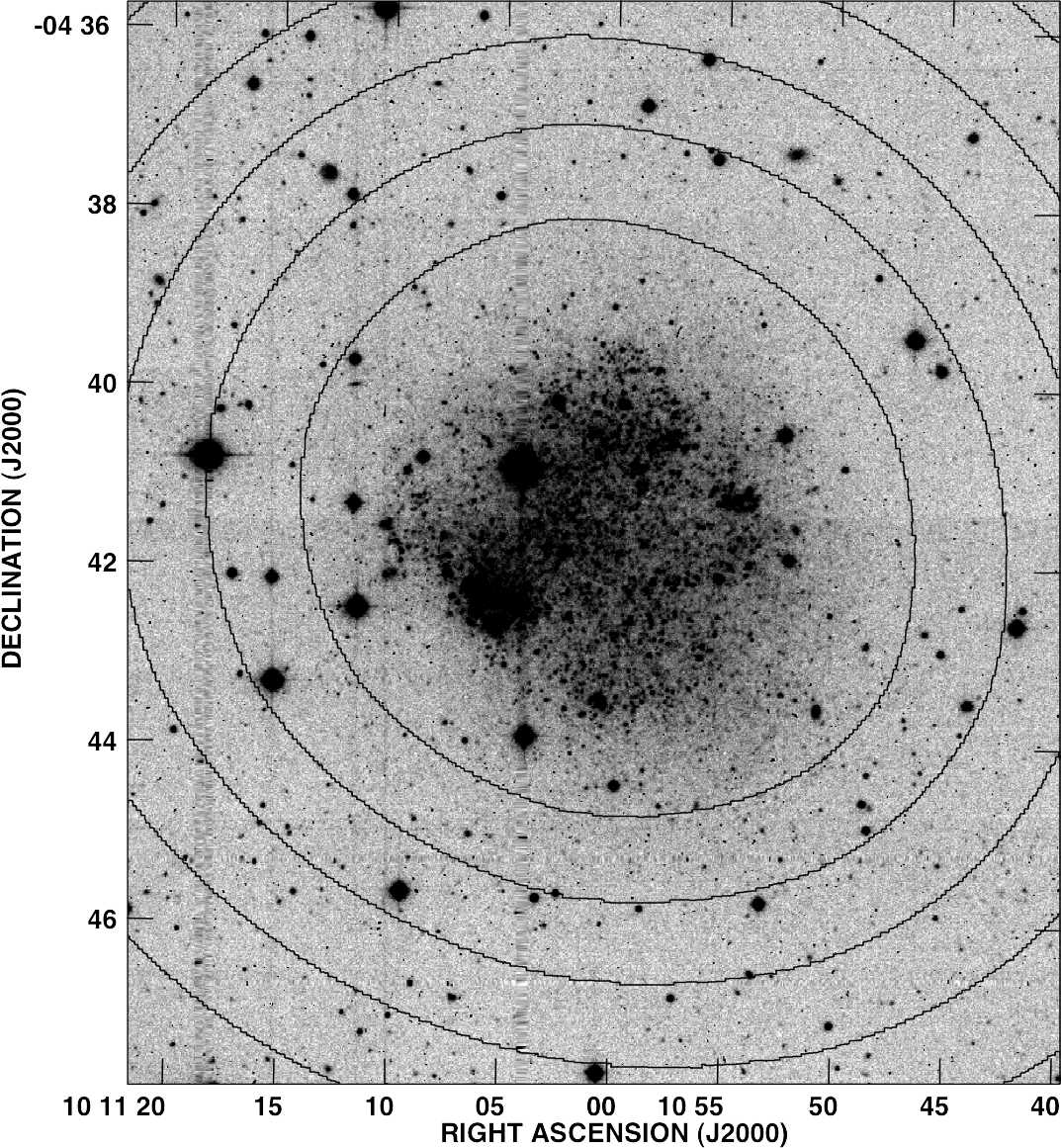}
\caption{$V$-band image of Sextans A with contours from the integrated \HI\ map superposed. 
Contours are the same as in Figure \ref{fig-sexam0}.
The $V$-band image is from Hunter \& Elmegreen (2006).
\label{fig-sexahionv}}
\end{figure}

Our integrated \HI\ map of Sextans A is shown in Figure \ref{fig-sexam0}
and contoured on a $V$-band image in Figure \ref{fig-sexahionv}.
Here we summed from 270.5 \protect\kms\
to 371.9 \protect\kms\ in the GBT data cube.
The central bright object is Sextans A, and we see no other structure in the map. 
Our 3$\sigma$ column density limit in Figure \ref{fig-sexam0}
is $2\times10^{18}$ \coldens. This is significantly better than the sensitivity limit of
$7.5\times10^{18}$ \coldens\ of Wilcots \& Hunter's (2002) VLA data and
the $10^{19}$ \coldens\ limit of Huchtmeier \et\ (1981).
In fact, this is only a little higher than the 3$\sigma$ limit of 
$1.6\times10^{18}$ \coldens\ achieved by Oosterloo \et\ (2007)
in ultra-deep mapping of the extended \HI\ halo around NGC 891.

Masses and \HI\ extents are given in Table \ref{tab-results}.
We calculate a total \HI\ mass of $8.6\times10^7$ M\solar. 
This is only 25\% more than the mass of $6.9\times10^7$
M\solar\ reported by Wilcots \& Hunter (2002), but 78\% of the
mass detected by Huchtmeier \et\ (1981), all corrected to the distance we are using.

\begin{figure}
\includegraphics[angle=0,width=1.0\textwidth]{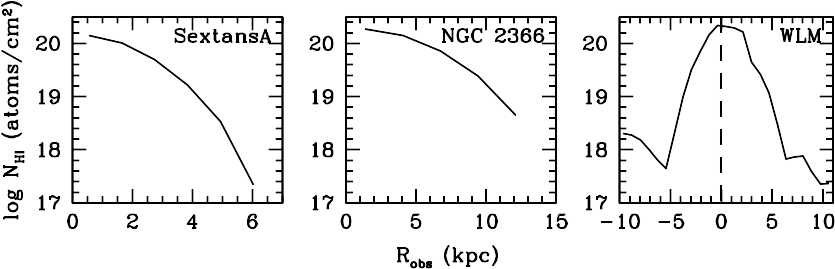}
\caption{For Sextans A and NGC 2366, integrated \protect\HI\ mass profiles measured 
on the moment zero maps in annuli 174\arcsec\ wide. Surface densities are 
corrected for inclination of the galaxy.
For WLM, we plot a north-south cut through the moment zero map, centered on the galaxy,
and averaged over 8.7\arcmin. The vertical line at zero marks the
center of the galaxy, and positive distance from that is to the north and negative to the south
away from the center.
$R_{obs}$ have not been corrected for convolution with the GBT 8.7\arcmin\ beam.
\label{fig-massprofiles}}
\end{figure}

It is immediately clear from Figure \ref{fig-sexam0} that there is
no large extended skirt of \HI\ around Sextans A. 
Figure \ref{fig-massprofiles} shows a column density profile
constructed from measuring the azimuthally-averaged surface density of the gas in annuli.
Because the beam-size is so large, the galaxy \protect\HI\ distribution
appears largely round. We used the position angle (P.A.) and inclination given in Table \ref{tab-obs},
and the column densities are corrected for inclination.
From the column density profile, we measure 
\HI\ around Sextans A to a diameter of 22.1\arcmin\
at a column density of $10^{19}$ \coldens. Corrected
for the 8.7\arcmin\ beam, this is 20.4\arcmin.
Hunter \& Wilcots (2002) reported a diameter of 18\arcmin\ 
at $7.5\times10^{18}$ \coldens\ from their VLA observations, 
and their surface density profile yields a diameter of 15.2\arcmin\ at the level of
$10^{19}$ \coldens\ that we are using for comparison here.
In short, we do not detect the extended disk of \HI\ that 
Huchtmeier \et\ reported in 1981. Wilcots and Hunter did not
detect this gas with the VLA, not because it is extremely smooth, but because
it is not there.

With Huchtmeier \et's \HI\ diameter of 54\arcmin, the ratio of
\HI\ diameter to optical diameter of Sextans A was extreme.
Figure 13 of Hunter (1997) compiles data on the \HI\ extents
of typical irregular galaxies. That figure is a histogram of the
ratio $R_{\rm HI}/R_{\rm Holm}$ of the radius of \HI\ at $10^{19}$ \coldens\ to
the optical extent as measured by the Holmberg radius,
defined at a photographic surface brightness
of 26.5 mag arcsec$^{-2}$. Most irregular galaxies, like most
spirals, have a ratio around 2. Only 6 systems were known at that
time to have $R_{\rm HI}/R_{\rm Holm}$ greater than 3, with the highest
being a ratio of 7. Sextans A was then thought to have the third highest
ratio with a value of 5.8, with the values of $R_{\rm HI}$ and $R_{\rm Holm}$
used at that time. 
With our new data, we find that Sextans A has an $R_{\rm HI}$ at
$10^{19}$ \coldens\ of 10.2\arcmin, and the ratio $R_{\rm HI}/R_{\rm Holm}$
is 3.3. 
(Our $R_{\rm Holm}$ are taken from Hunter \& Elmgreen [2006] and given in Table \ref{tab-sample}). 
This now puts Sextans A at the high end of the distribution of
the bulk of other irregular galaxies and an \HI\ extent that is not particularly
abnormal.

\begin{figure}
\includegraphics[angle=0,width=1.0\textwidth]{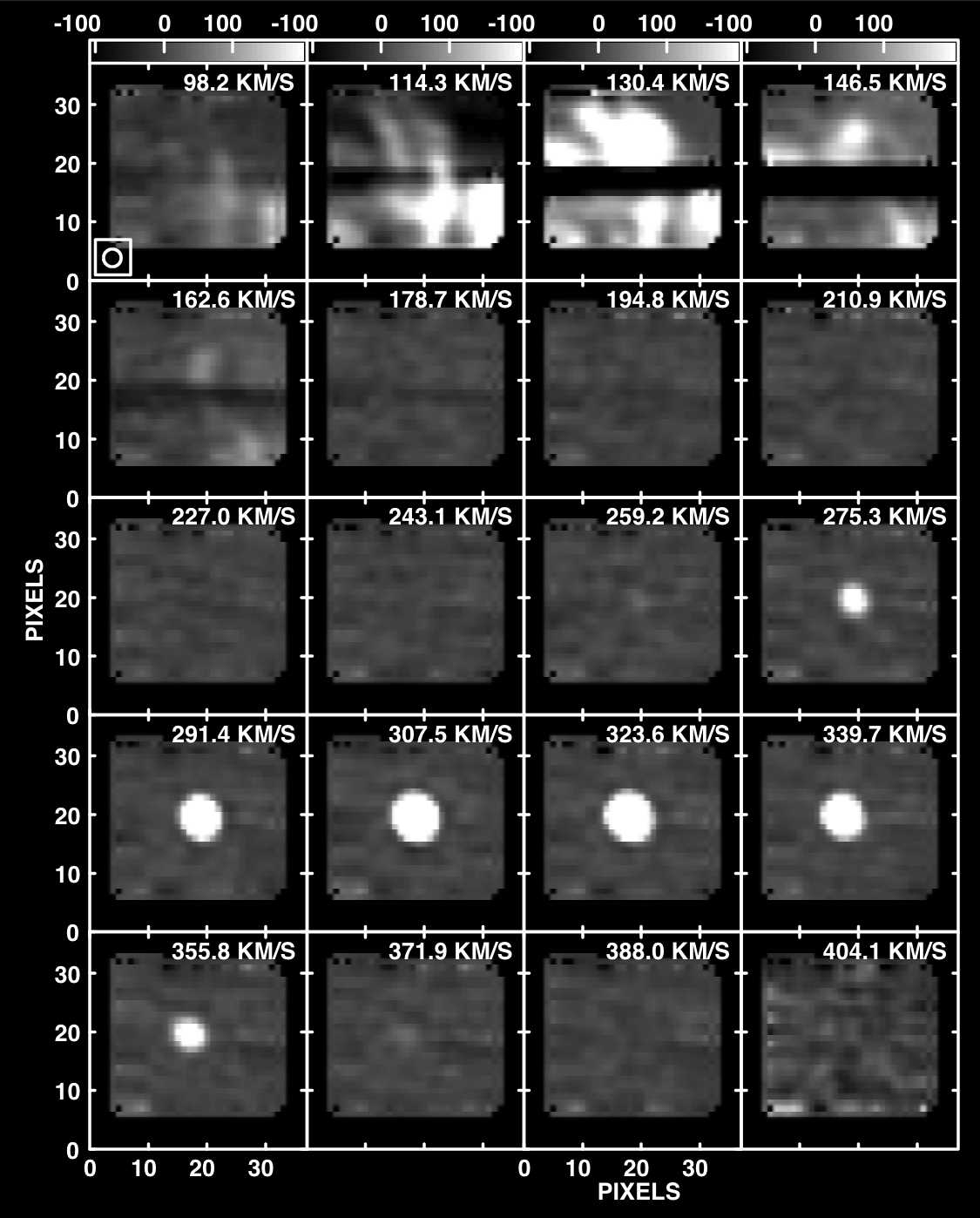}
\caption{Sextans A channel maps composed from the GBT data cube to match
as closely as possible the central velocity and channel spacing
of channel maps shown by Huchtmeier et al. (1981) in their Figure 3c. 
We have summed over 10 of our channels (16.1 \protect\kms) to produce each of these channel
maps.
The circle in the bottom left corner of the top left panel is the FWHM of the beam: 8.72\arcmin.
The gray scale along the top is flux in units of K beam$^{-1}$ pixel$^{-1}$.
\label{fig-sexakntr}}
\end{figure}

Where could the discrepancy between the Huchtmeier \et\ (1981)
Effelsberg observations and the GBT observations lie?
We have replicated as closely as possible the channel maps
of Sextans A given in Figure 3c of Huchtmeier \et\ Since their
velocity resolution is lower than ours, this involved summing
over the appropriate channels in the GBT data cube. These channel maps
are shown in Figure \ref{fig-sexakntr}. Each map
is the average of 10 channels and is centered as closely
as possible on the same velocities as Huchtmeier \et's channel maps.
Figure \ref{fig-sexakntr} looks different than Figure 3c
of Huchtmeier \et\ The Effelsberg data detect extended \HI\ that
does not appear in our maps and the shape of Sextans A at
the lowest column densities differs. 

There are significant differences between the GBT observations and
those done at Effelsberg that may contribute to differences
in the maps. First, although the beam sizes are nearly
the same, Huchtmeier \et\ (1981) mapped Sextans A every beam-size
(9\arcmin), whereas we mapped every 1/3 of a beam (2.9\arcmin).
Thus, our map is better sampled.
Second, our signal-to-noise is higher, and the extended features in
the Huchtmeier \et\ maps occur at low signal-to-noise.
Third, although the GBT does have side-lobe problems (Robishaw \& Heiles 2009), 
they are different from those at Effelsberg and we
believe that the Effelsberg map was confusing
Galactic emission with emission from Sextans A.
We detect Galactic emission at lower velocities, offset about 120 \kms.

\subsubsection{WLM}

\begin{figure}
\includegraphics[angle=0,scale=0.9,width=1.0\textwidth]{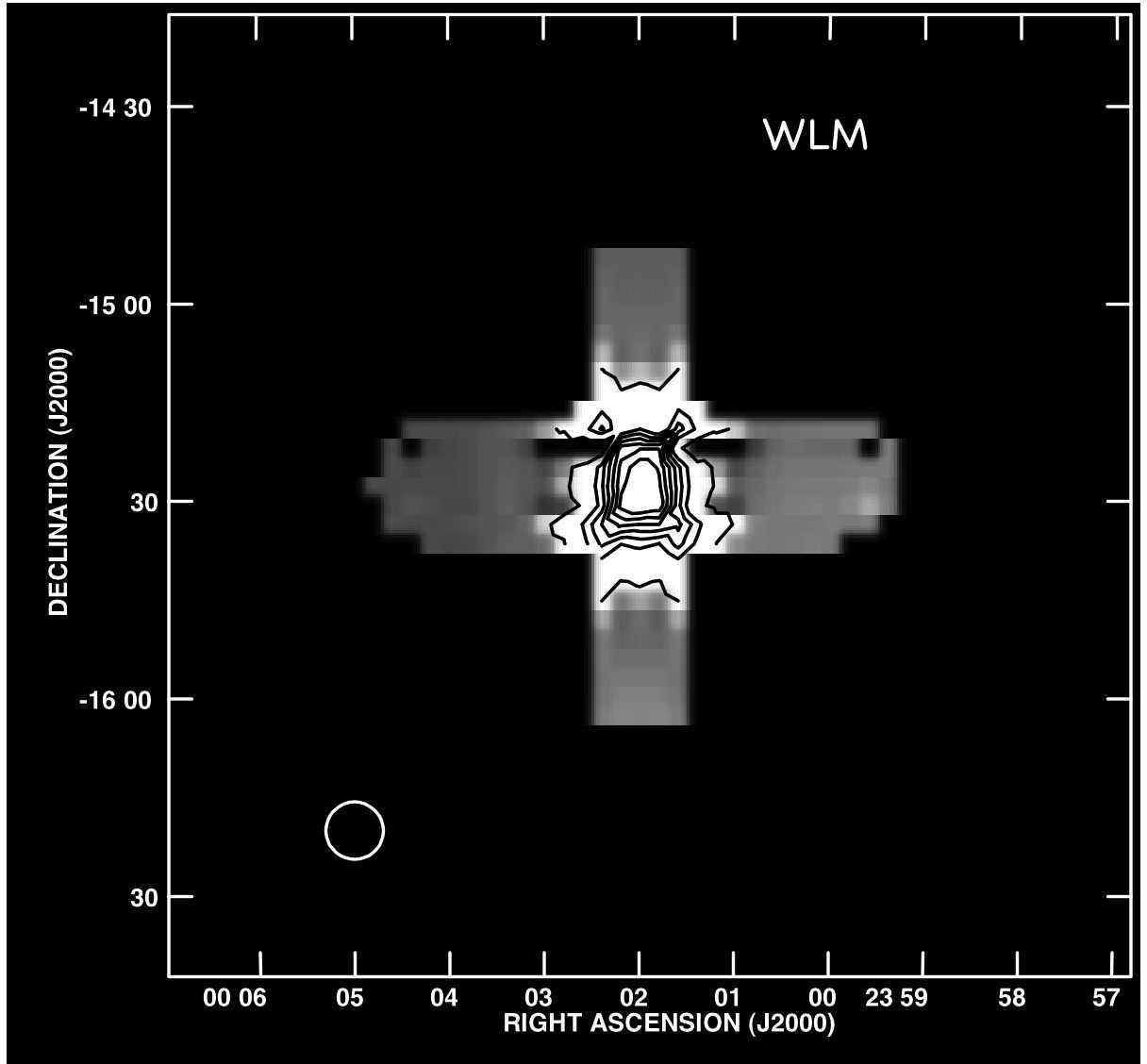}
\caption{Integrated \HI\ map of WLM summed from -179 \protect\kms\
to -71 \protect\kms\ in the GBT data cube. The image is displayed
to bring out the fainter outlying structure.
The circle in the bottom left corner is the FWHM of the beam: 8.72\arcmin\ diameter.
Map pixels are 2.9\arcmin, 1/3 of a beam-size.
Contours are 15.3, 100, 140, 180, 220, 260, and 300 K beam$^{-1}$ km s$^{-1}$ pixel$^{-1}$.
The lowest contour corresponds to a column density of $10^{19}$ cm$^{-2}$, the isodensity contour
to which we have measured the HI diameters.
The incomplete nature of the map results in imaging artifacts.
\label{fig-wlmm0}}
\end{figure}

\begin{figure}
\includegraphics[angle=-90, scale=0.85]{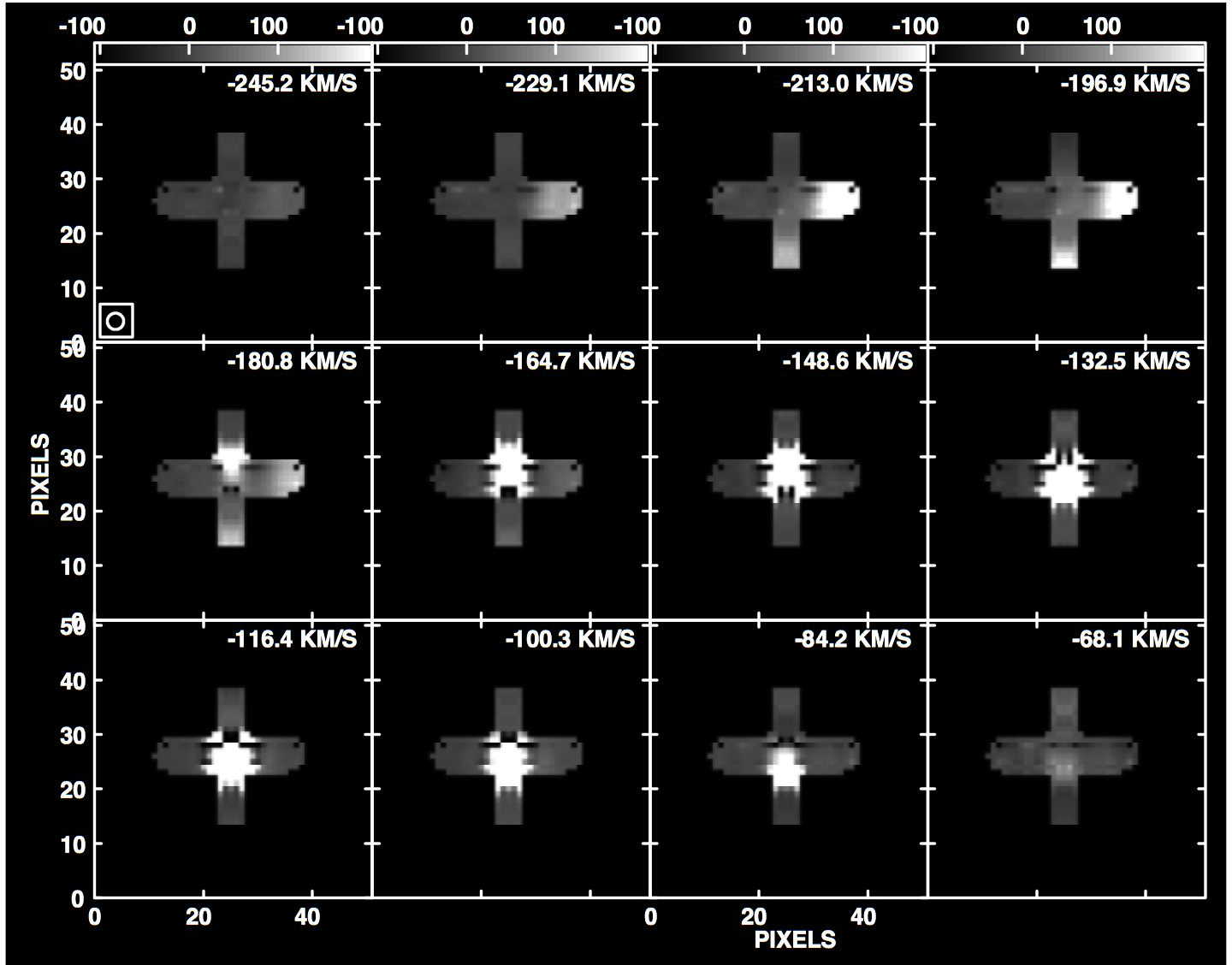}
\caption{$V$-band image of WLM with contours from the integrated \HI\ map superposed. 
Contours are the same as in Figure \ref{fig-wlmm0}.
The incomplete nature of the map results in imaging artifacts.
The $V$-band image is from Hunter \& Elmegreen (2006).
\label{fig-wlmhionv}}
\end{figure}

We detected \HI\ emission from WLM between -179 \kms\ to -71 \kms.
Similarly, Kepley \et\ (2007) detected emission from -170 \kms\ to -72 \kms
and Huchtmeier \et\ (1981) from -179 \kms\ to -64 \kms.
Our integrated \HI\ map of WLM is shown in Figure \ref{fig-wlmm0}
and contours are superposed on a $V$-band image in Figure \ref{fig-wlmhionv}.
Because of limited telescope time, our map consists of a strip through the center along
the north-south direction and along the east-west direction. 
This was done in order to look for extended emission along two diameters.
Thus, the map is not complete and imaging artifacts are obvious.
However, no extended skirt of \HI\ is seen along the two map directions. 
Our 3$\sigma$ column density limit in Figure \ref{fig-wlmm0} is $9\times10^{17}$ \coldens. 

Masses and \HI\ extents are given in Table \ref{tab-results}.
We calculate a total \HI\ mass of $1.1\times10^8$ M\solar. 
This is 1.6$\times$ more than the mass of $7.0\times10^7$
M\solar\ reported by Kepley \et\ (2007), 1.9$\times$ the
mass of $5.9\times10^7$ M\solar\ detected by Huchtmeier \et\ (1981),
3 times higher than the mass of Jackson \et\ (2004), and
60\% higher than Barnes \& De Blok's (2004) mass
---all corrected to our choice of distance for WLM.
However, because of the structure of our map, our total mass is uncertain.

For WLM, rather than constructing an azimuthally-averaged column density profile
since our map is incomplete, we made a simple north-south cut through
the center of the galaxy. This is close to the major axis of the galaxy.
This cut sums east-west over 3 pixels (8.7\arcmin),
and is shown in Figure \ref{fig-massprofiles}. The distance from 
the center of the galaxy is positive for the north direction and negative for south.
From this cut, we see that  a column density of $10^{19}$ \coldens\ occurs
at a radius of 12.9\arcmin\ to the south and 16.4\arcmin\ to the north of the galaxy center.
The average diameter is 29.3\arcmin. Corrected
for the 8.7\arcmin\ beam, this is 28\arcmin, which is comparable
to the diameter of 29.3\arcmin\ from the VLA observations of Kepley \et\ (2007).
Huchtmeier \et\ (1981), on the other hand, measure a diameter of 45\arcmin. 
Our diameter gives an $R_{\rm HI}/R_{\rm Holm}$ ratio of 2.4 (using the $R_{\rm Holm}$ in 
Table \ref{tab-sample}), which makes it normal compared to other dwarfs.

\begin{figure}
\includegraphics[angle=0,width=1.0\textwidth]{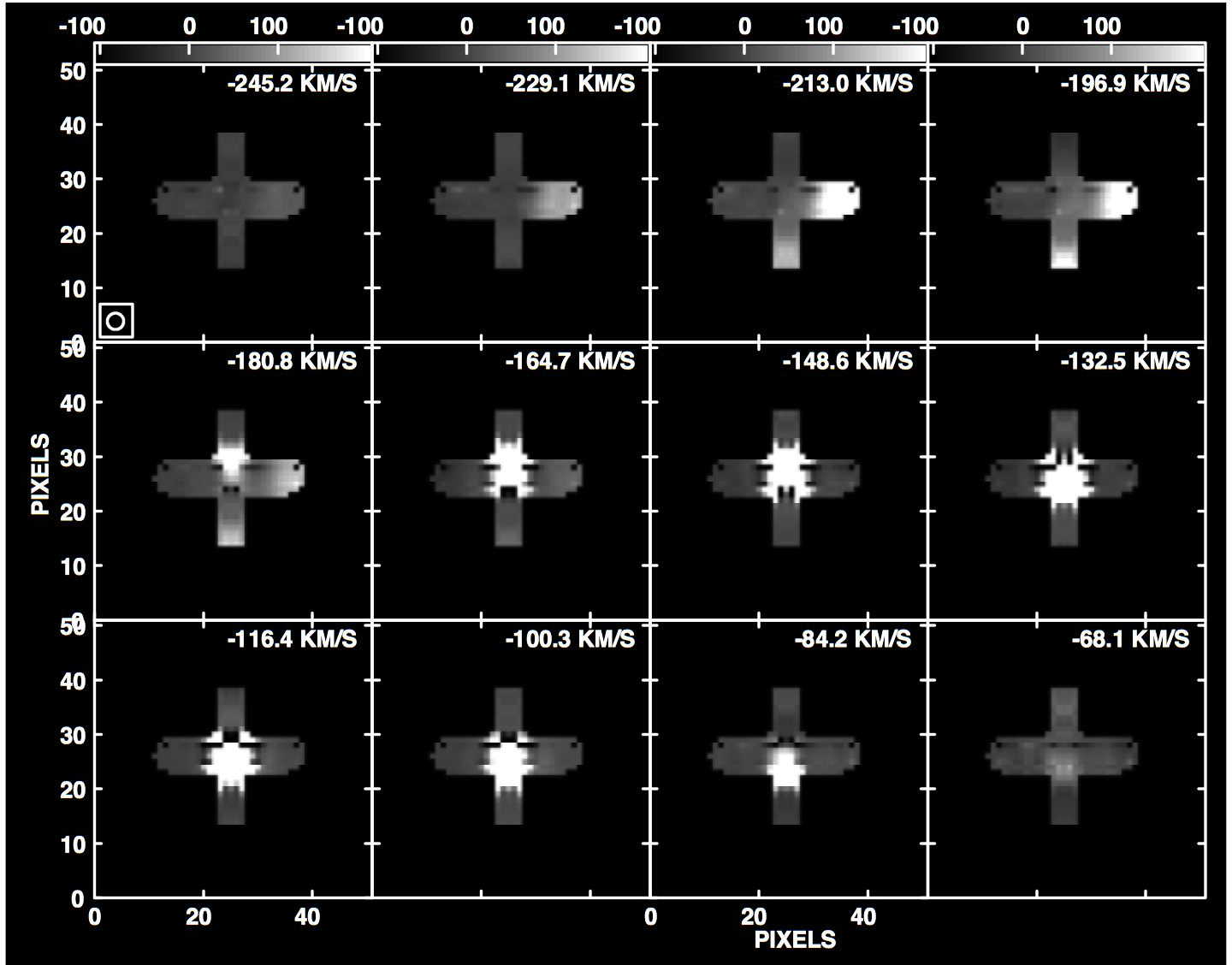}
\caption{WLM channel maps composed from the GBT data cube to match
as closely as possible the central velocity and channel spacing
of channel maps shown by Huchtmeier et al. (1981) in their Figure 3d. 
We have summed over 10 of our channels (16.1 \protect\kms) to produce each of these channel
maps.
The circle in the bottom left corner of the top left panel is the FWHM of the beam: 8.72\arcmin.
The gray scale along the top is flux in units of K beam$^{-1}$ pixel$^{-1}$.
\label{fig-wlmkntr}}
\end{figure}

We have replicated the channel maps
of WLM given in Figure 3d of Huchtmeier \et\ (1981) by summing
over 10 channels in the GBT data. These channel maps
are shown in Figure \ref{fig-wlmkntr}. 
Huchtmeier \et\ detect unusually extended emission at
velocities of -163 \kms\ to -229 \kms. Figure \ref{fig-intprofiles}
shows that we have not detected emission at those velocities.
However, we do detect Galactic emission between +50 \kms\ and -50 \kms.
WLM is also known to be near the Magellanic Stream (Putman \et\ 2003).

\subsubsection{NGC 2366}

\begin{figure}
\includegraphics[angle=0,scale=0.9,width=1.0\textwidth]{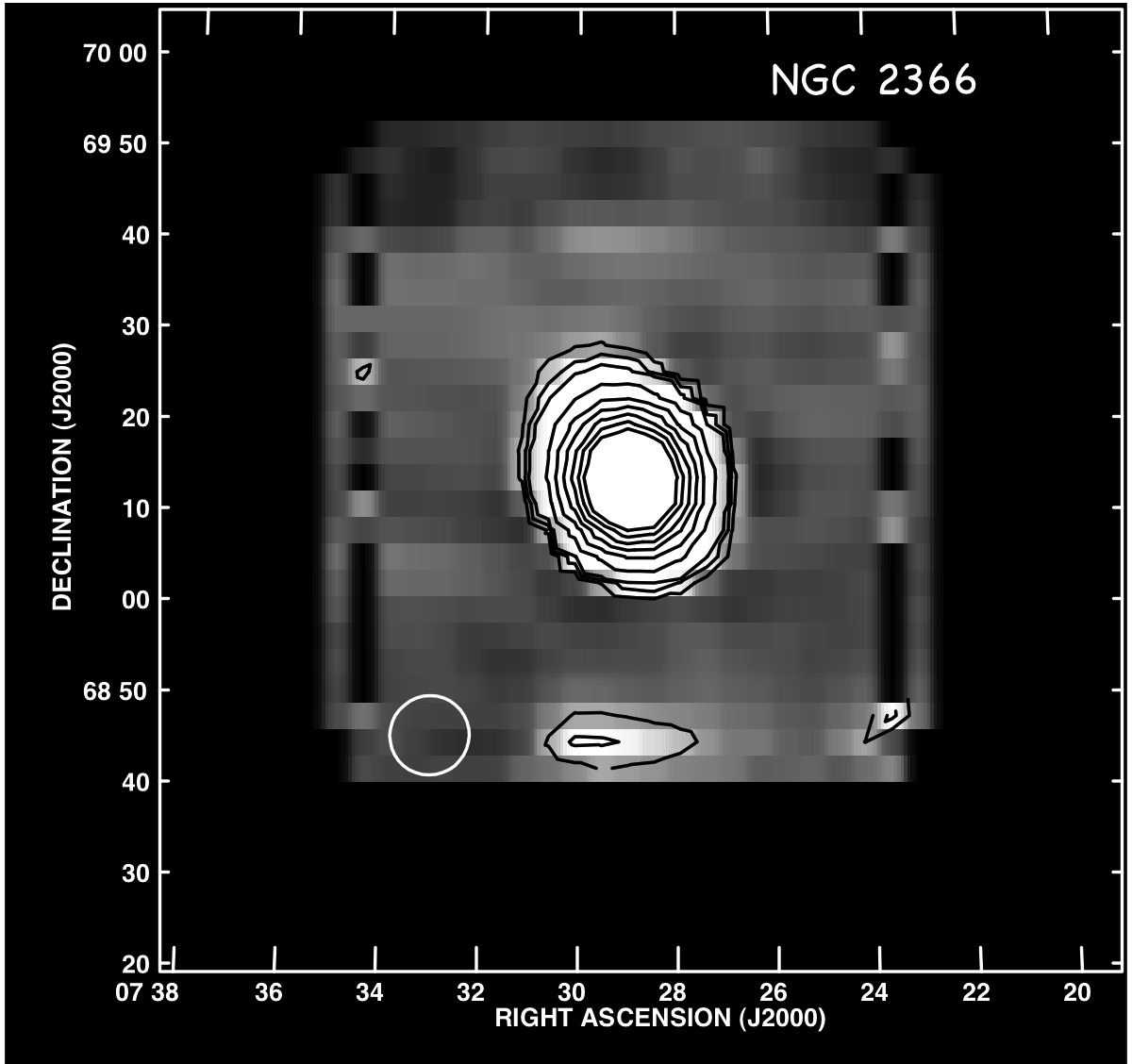}
\caption{Integrated \HI\ map of NGC 2366 summed from 30.5 \protect\kms\
to 172.1 \protect\kms\ in the GBT data cube. The image is displayed
to bring out the fainter outlying structure.
The circle in the bottom left corner is the FWHM of the beam: 8.72\arcmin\ diameter.
Map pixels are 2.9\arcmin, 1/3 of a beam-size.
Contours are 5, 10, 15.3, 40, 70, 100, 130, 160, and 190 K beam$^{-1}$ km s$^{-1}$ pixel$^{-1}$.
The third contour corresponds to a column density of $10^{19}$ cm$^{-2}$, the isodensity contour
to which we have measured the HI diameters.
Examination of the data cube suggests that the emission along the bottom and right edges of the map
is noise. 
\label{fig-n2366m0}}
\end{figure}

\begin{figure}
\includegraphics[angle=0,scale=0.9,width=1.0\textwidth]{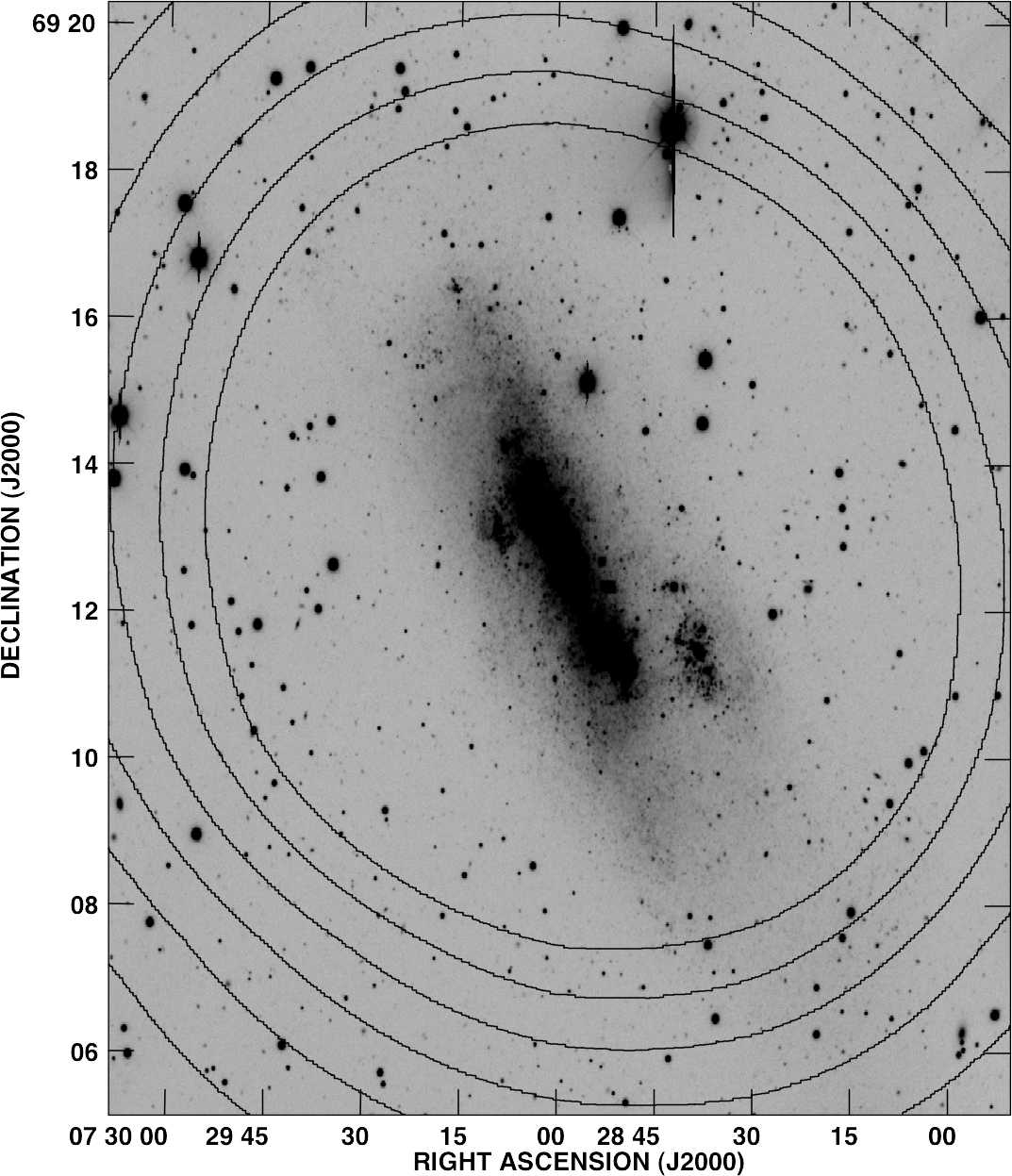}
\caption{$V$-band image of NGC 2366 with contours from the integrated \HI\ map superposed. 
Contours are the same as in Figure \ref{fig-n2366m0}.
The $V$-band image is from Hunter \& Elmegreen (2006).
\label{fig-n2366hionv}}
\end{figure}

We detected \HI\ emission from NGC 2366 from 30 \kms\ to 172 \kms, 
very close to what Hunter \et\ (2001) found.
Our integrated \HI\ map of NGC 2366 is shown in Figure \ref{fig-n2366m0},
and contours are plotted on the $V$-band image in Figure \ref{fig-n2366hionv}.
No extended skirt of \HI\ is seen in this map. 
Our 3$\sigma$ column density limit in Figure \ref{fig-wlmm0} is $2.5\times10^{18}$ \coldens. 

Masses and \HI\ extents are given in Table \ref{tab-results}.
We calculate a total \HI\ mass of $7.4\times10^8$ M\solar. 
This is 7\% higher than the mass of $6.9\times10^8$
M\solar\ reported by Hunter \et\ (2001), 4\% higher than the mass
detected by Huchtmeier \et\ (1981), 60\% higher than Wevers \et's (1986) mass, 
14\% higher than Swaters \et's (2002) mass, and 30\% higher than Walter \et's (2008) mass,
all corrected to the distance we are using.

Figure \ref{fig-massprofiles} shows a column density profile
constructed from measuring the azimuthally-averaged surface density of the gas in annuli
(see Table \ref{tab-obs}).
We detect \HI\ at a column density of $10^{19}$ \coldens\ around NGC 2366 at
a diameter of 23.4\arcmin. Corrected
for the 8.7\arcmin\ beam, this is an extent of 21.7\arcmin. 
This is similar to the extent reported by Hunter \et\ (2001)
from VLA observations, but less than the 30\arcmin\ diameter
reported by Huchtmeier \et\ (1981).
Hunter \et\ report an \HI\ diameter of 23.4\arcmin\ at $1.7\times10^{19}$ \coldens,
and their surface density profile yields 25\arcmin\ at the level of $10^{19}$ \coldens\
that we are using for comparison here.
Our \HI\ extent gives a ratio of $R_{\rm HI}/R_{\rm Holm}$ of 2.3, 
which is normal compared to other dwarfs.

\begin{figure}
\includegraphics[angle=0,width=1.0\textwidth]{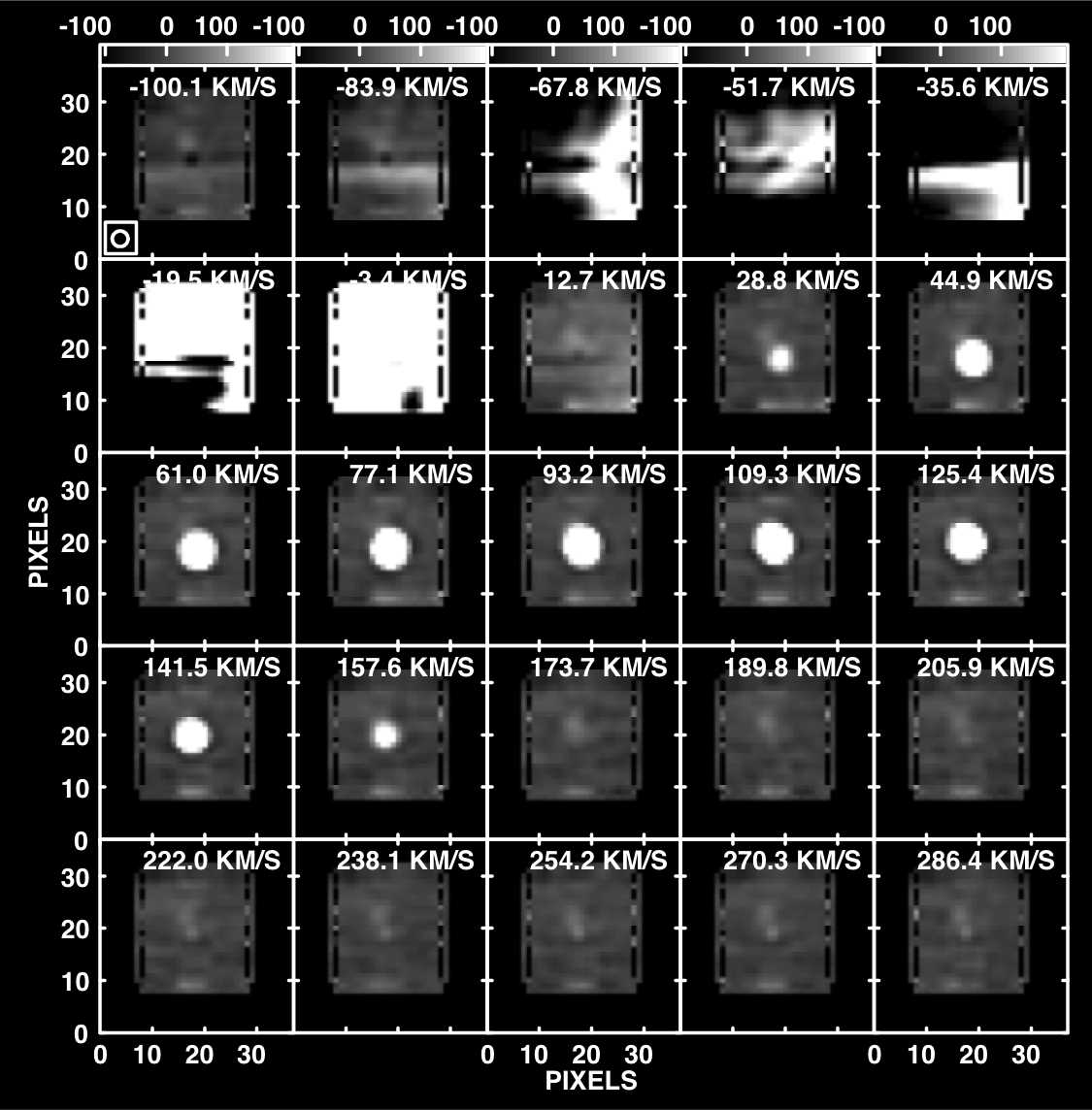}
\caption{NGC 2366 channel maps composed from the GBT data cube. 
We have summed over 10 of our channels (16.1 \protect\kms) to produce each of these channel
maps.
The circle in the bottom left corner of the top left panel is the FWHM of the beam: 8.72\arcmin.
The gray scale along the top is flux in units of K beam$^{-1}$ pixel$^{-1}$.
\label{fig-n2366kntr}}
\end{figure}

We have produced channel maps
of NGC 2366 similar to those produced for Sextans A and WLM even
though there is no Huchtmeier \et\ (1981) figure for comparison.
We summed over 10 channels in the GBT data. These channel maps
are shown in Figure \ref{fig-n2366kntr}. 
We detect Galactic emission from -100 \kms\ to about 25 \kms.

\subsection{The \HI\ Fall-Off}

What can we learn about the distribution of \HI\ in the outer
part of these galaxies?
Figure \ref{fig-massprofiles} shows the column density \HI\ profiles.
For WLM, where we have a north-south cut, the distribution
is Gaussian in shape and shows a rather rapid drop to 
our detection limit.
For Sextans A and NGC 2366, where we have azimuthally-averaged
surface densities, we also see a rather rapid drop at large radii (the logarithm
of the surface density is plotted).
Figure 4 of Wilcots \& Hunter (2002) also gives an \HI\ surface density
profile of Sextans A along with those of several other irregular
galaxies. The outer \HI\ of Sextans A is seen in that figure
to drop off in a fashion that is
similar to several other irregular galaxies and a little more
or less rapidly than others.
We can say that there is no abrupt truncation
in the \HI\ at a limit $\geq10^{18}$ \coldens\ in Sextans A and at a limit $\geq10^{19}$ \coldens\ in
NGC 2366 and WLM.

Observations of low-z absorbers imply that galaxies could extend
to as low a column density as $10^{14}$ \coldens\ (e.g.
Penton, Stocke, \& Shull 2002) if Ly$\alpha$ absorbers are associated with individual
galaxies as we know them today.
However, theoretical arguments suggest that all galaxies should have sharp
edges at the same column density due to the metagalactic 
UV background radiation (Maloney 1993).
Some studies have found abrupt truncations and others have not.
Van Gorkom (1993) reports that the \HI\ in the spiral NGC 3198 is
sharply truncated at $2\times10^{19}$ \coldens, and
Corbelli \et\ (1989) see a steep fall-off to $2\times10^{18}$
\coldens\ in M33. On the other hand, neither Walsh \et\ (1997) 
nor Carignan \& Purton (1998) found truncations
to $1\times10^{19}$ \coldens\ around NGC 289 and DDO 154,
and Portas \et\ (2009) find that the \HI\ disks in spirals drop off
similarly to their detection limit.
Our observations do not support disk truncation.

Limits to the \HI\ edges are expected to come from internal
sources of ionizing photons as well as from the metagalactic UV background
radiation. 
A warp in the gas disk can allow ionizing flux escaping from the galaxy to dominate
ionization of the outer \HI. 
For example, Bland-Hawthorn \et\ (1997) found ionized gas beyond
the \HI\ disk in NGC 253. Yet the emission measures of \ha\ and [NII]
are too high for the source of ionization to be the metagalactic UV background.
Instead, they argue that the source of ionization is hot young stars in the inner 
regions of the galaxy that see the warped outer \HI\ disk. 
Certainly, all three of our galaxies have modest on-going star formation.
The ionizing background radiation would produce
an edge that is independent of azimuth while structure from photons escaping from within
would depend on the distribution of the galaxy's gas disk (Maloney 1993). 
We do not see evidence for an azimuthal variation in the outer \HI\ in our galaxies.

On the other hand, in a study of nearby spiral galaxy gas disks, 
Portas \et\ (2009) found that the gas density drops sharply at large radii.
They argue that the fall-off of \HI\ begins in spiral disks
at too high a column density ($10^{20}$ \coldens) to be due to the metagalactic UV background. 
Furthermore, the shape of the fall-off is similar among a wide variety of spiral galaxies.
They argue instead that the shape of the fall-off of the gas density with radius is more likely 
due to how gas disks form, reflecting their dark matter haloes and gas accretion history.
Our data do not contradict this suggestion.

Our results likely have interesting implications for the nature of the intergalactic medium on
the outskirts of the Local Group.  Sextans A is certainly on the outskirts and WLM might be
as well.  The fact that we are not seeing a truncation may imply that these galaxies have yet
to encounter the intragroup radiation field in the Local Group and certainly are not yet subject
to any of the gas removal processes associated with the infall of galaxies into groups.

\section{Summary} \label{sec-summary}

With GBT single-dish pointings sampling in 2.9\arcmin\ steps,
we have constructed deep large field-of-view maps of the \HI\
emission centered on three dwarf irregular galaxies.
We do not detect the extended skirt of \HI\ reported by Huchtmeier \et\ (1981)
in these galaxies,
and conclude that Galactic emission in the field of view and
side-lobe emission may have led to confusion with the sources.
We measure new diameters for the \HI\ extents at
$10^{19}$ \coldens.
With these new dimensions, the ratios of \HI\ extent to optical extent 
are normal for dwarf irregular galaxies.
In addition, there is no evidence for a truncation in the \HI\ distribution
of these galaxies $\geq10^{19}$ \coldens.

\acknowledgments

We wish to thank Dr. Glen Langston, NRAO staff member at Green Bank, for assistance
with implementing our observing program and in understanding how
to reduce the data with aips$++$.
We also are grateful for assistance from Megan Johnson with understanding how
to reduce these data in GBTIDL.
ECB would like to thank Dr. Kathy Eastwood and the 2003
Research Experience for Undergraduates program at Northern Arizona
University which was funded by the National Science Foundation under grant 9988007.
FZ acknowledges the 2011 MIT Field Camp at Lowell Observatory and Dr. Amanda
Bosh for organizing that.
Funding for this work was also provided
by the National Science Foundation through grant AST-0204922 to DAH.

Facilities: \facility{Green Bank Telescope}

\clearpage

\clearpage

\clearpage

\clearpage

\clearpage

\clearpage

\clearpage

\clearpage

\clearpage

\clearpage

\clearpage

\clearpage

\begin{deluxetable}{llccccc}
\tabletypesize{\scriptsize}
\rotate
\tablecaption{Galaxy Sample \label{tab-sample}}
\tablehead{
\colhead{} & \colhead{} & \colhead{D\tablenotemark{b}} & \colhead{} & \colhead{}
& \colhead{$R_{Holm}$\tablenotemark{d}} 
& \colhead{$R_D$\tablenotemark{e}} \\
\colhead{Galaxy} & \colhead{Other Names\tablenotemark{a}}
& \colhead{(Mpc)}
& \colhead{$M_V$\tablenotemark{b}}
& \colhead{E(B$-$V)$_f$\tablenotemark{c}}
& \colhead{(kpc)} & \colhead{(kpc)}}
\startdata
Sextans A  & PGC 29653, UGCA 205,DDO 75                         &  1.3 & $-$13.9 & 0.02 & 1.17 & 0.22 \\
WLM           & PGC 143, UGCA 444, DDO 221, A2359$-$15  &  1.0 & $-$14.4 & 0.02 & 1.69 & 0.57 \\
NGC 2366 & PGC 21102, UGC 3851, DDO 42                        &   3.2 & $-$16.7 & 0.04 & 4.40 & 1.28 \\
\enddata\tablenotetext{a}{Selected alternate identifications obtained from NED.}
\tablenotetext{b}{References for distances and $M_V$ can be found in Hunter \& Elmegreen (2006).}
\tablenotetext{c}{Foreground Milky Way reddening from Burstein \& Heiles (1984).}
\tablenotetext{d}{Holmberg radius, in kpc, measured at a $B$-band surface brightness of
26.66 mag arcsec$^{-2}$. These values are taken from Hunter \& Elmegreen (2006).}
\tablenotetext{e}{Disk scale length, in kpc, determined from $V$-band surface photometry, 
from Hunter \& Elmegreen (2006).}
\end{deluxetable}

\clearpage

\begin{deluxetable}{lcccccc}
\tabletypesize{\scriptsize}
\rotate
\tablecaption{Observation and Analysis Parameters \label{tab-obs}}
\tablehead{
\colhead{} & \colhead{V$_{cen}$} 
& \colhead{Map center} 
& \colhead{} 
& \colhead{Final Map\tablenotemark{a}}
& \colhead{P.A.\tablenotemark{b}}
& \colhead{Incl\tablenotemark{b}} \\
\colhead{Galaxy} & \colhead{(km s$^{-1}$)}
& \colhead{(RA,DEC)}
& \colhead{Beam pointings\tablenotemark{a}}
& \colhead{(arcmin$\times$arcmin)}
& \colhead{(deg)}
& \colhead{(deg)} 
}
\startdata
Sextans A  &      324 & 10$^{\rm h}$ 11$^{\rm m}$ 01.3$^{\rm s}$, $-$4\arcdeg\ 41\arcmin\ 15\arcsec  & 25$\times$24 & 87$\times$78 &   47 & 14  \\
WLM          & $-$116 & 0$^{\rm h}$ 01$^{\rm m}$ 59.2$^{\rm s}$, $-$15\arcdeg\ 27\arcmin\ 41\arcsec  & 25$\times$25 & 72$\times$72 & 174 & 0  \\
NGC 2366 &      100 & 7$^{\rm h}$ 29$^{\rm m}$ 00.1$^{\rm s}$, $+$69\arcdeg\ 13\arcmin\ 18\arcsec & 18$\times$20 & 55$\times$72 &   46 & 28  \\
\enddata
\tablenotetext{a}{The map of WLM consists of an E-W strip 3 pointings wide (one offset 1/3 beam width to the north
and one offset 6\arcsec\ to the south) and a single N-S strip through the galaxy center, 
with the lengths given here.}
\tablenotetext{b}{P.A. is from the optical morphology (Hunter \& Elmegreen 2006), 
and the inclination is from the apparent minor-to-major
axis ratio of the GBT \protect\HI\ distribution. These are used for the \protect\HI\ surface density profiles.}
\end{deluxetable}

\clearpage

\begin{deluxetable}{lcccccccccc}
\tabletypesize{\scriptsize}
\rotate
\setlength{\tabcolsep}{0.02in}
\tablecaption{\protect\HI\ Masses and Extents \label{tab-results}}
\tablehead{
\colhead{} 
& \colhead{} 
& \colhead{} 
& \colhead{} 
& \colhead{} 
& \multicolumn{6}{c}{\protect\HI\ extent at $10^{19}$ \protect\coldens\tablenotemark{c}} \\

\cline{6-11}

\colhead{} 
& \multicolumn{3}{c}{Integrated \protect\HI\ mass (M\protect\solar)\tablenotemark{a}}
& \colhead{}
& \multicolumn{2}{c}{This study}
& \colhead{}
& \colhead{Literature\tablenotemark{b}}
& \colhead{}
&  \colhead{Huchtmeier et al. 1981} \\
\cline{2-4}
\cline{6-7}

\colhead{Galaxy} 
& \colhead{This study}
& \colhead{Literature\tablenotemark{b}}
& \colhead{Huchtmeier et al. 1981}
& \colhead{}
& \colhead{$R_{\rm HI}$ (\arcmin)}
& \colhead{$R_{\rm HI}/R_{\rm Holm}$\tablenotemark{d}} 
& \colhead{}
& \colhead{$R_{\rm HI}$ (\arcmin)}
& \colhead{} 
& \colhead{$R_{\rm HI}$ (\arcmin)}
}
\startdata
Sextans A  & $8.6\times10^7$ & $6.9\times10^7$    & $1.1\times10^8$ & & 10.2 & 3.3 & & 7.6 & & 27  \\
WLM           & $1.1\times10^8$ & $7.0\times10^7$  & $5.9\times10^7$ & & 14.0 & 2.4 & & 14.7  & & 22.5 \\
NGC 2366 & $7.4\times10^8$ & $6.9\times10^8$ & $7.1\times10^8$ & & 10.8 & 2.3 & & 12.5 & & 15  \\
\enddata
\tablenotetext{a}{Masses are corrected from the original values given in the literature
to the distances given in Table \protect\ref{tab-sample}.}
\tablenotetext{b}{References: Sextans A---Wilcots \& Hunter (2002),
NGC 2366---Hunter et al. (2001), WLM---Kepley et al. (2007).}
\tablenotetext{c}{We report here extents of \protect\HI\  to $10^{19}$ \protect\coldens. For Sextans A
and NGC 2366, the extents reported in the literature are for different isodensity levels:
Sextans A has an $R_{\rm HI}$ of 9\protect\arcmin\ at $8\times10^{18}$ \protect\coldens,
and NGC 2366 has an $R_{\rm HI}$ of 11.7\protect\arcmin\ at $1.7\times10^{19}$ \protect\coldens.
For these galaxies, we have used the surface density profiles from the original studies to 
determine the extent at $10^{19}$ \protect\coldens, and those values are reported in this Table.
}
\tablenotetext{d}{The $R_{\rm Holm}$ used here are taken from Hunter \& Elmegreen (2006)
and given in Table \protect\ref{tab-sample}.}
\end{deluxetable}

\end{document}